\documentclass{article}
\begin{document}
\hfill{\bf SB/F/01-289}\medskip\\
\begin{center}
{\textbf{\Large{Satellite potentials for hypergeometric Natanzon potentials}}}
\end{center}
\vspace{.5in}
\begin{center}
\textbf{S Codriansky$^{a}$\footnote{codrians@reacciun.ve} and S Salam\'{o}$^{b}$\footnote{ssalamo@fis.usb.ve}}
\end{center}
\begin{center}
\textit{$^{a}$Departamento de Matem\'{a}ticas y F\'{\i}sica, Instituto Pedag\'{o}%
gico de Caracas}\\
\textit{Universidad Pedag\'{o}gica Experimental Libertador}\\
\textit{$^{b}$Universidad Sim\'{o}n Bol\'{\i}var,Departamento de F\'{\i}sica}
\textit{Apartado Postal 89000, Caracas, Venezuela}
\end{center}
{PACS 03.65.Fd - Algebraic methods}

\begin{abstract}
As a result of the $so(2,1)$ treatment of the hypergeometric Natanzon
potentials a set of potentials related to a given one is determined. The set
arises as a result of the action of the $so(2,1)$ generators on the carrier
space of an irreducible representation.
\end{abstract}

\section{Introduction}

In \cite{corsa} an $so(2,1)$ description of a given hypergeometric Natanzon %
\cite{nata} potential ${V_{N}}$ was presented. The study of the discrete
spectrum led to the result that three parameters (called group parameters)
are required to completely describe the eigenstates. Two of these parameters
correspond to labels of states in a particular irreducible representation
(irrep) of $so(2,1)$: the eigenvalues $q$ of the Casimir operator and $m$ of
the compact generator. The third parameter, $p$, labels a particular set of $%
so(2,1)$ generators. It is found that in most cases there is one eigenstate
for each value of $p$ and as a result different sets of $so(2,1)$
generators, each for a given (allowed) value of $p$, are necessary to describe
the eigenfunctions of ${V_{N}}$. It is unusual that all the eigenstates of ${%
V_{N}}$ belong to a single irrep so that if attention is fixed in a
particular irrep the states connected by the $so(2,1)$ generators are, in
the majority of cases, associated with different Natanzon potentials. It is
this fact that suggests the definition of a satellite potential. The
construction of the set of satellite potentials associated to ${V_{N}}$ is
the content of this paper. The term satellite potential is used in the
present paper as in \cite{mesa}. \medskip \newline 
\indent The action of the $so(2,1)$ generators defines a set of potentials
which share the property that the eigenstates connected by them have the
same $p$ and $q$ but correspond, in general, to different energy
eigenvalues. If the energy eigenvalues of the states connected by the $%
so(2,1)$ generators are the same the following question naturally arises:
are the $so(2,1)$ generators related to the operators in supersymmetric
quantum mechanics (SUSYQM) \cite{cuper1} This is not so for the shape
invariant potentials ${V_{S}}$ \cite{gede} considered in this paper, as is
shown below. It is also proven that the state reached by the action of the $%
so(2,1)$ generators belongs to a Natanzon potential. This implies that the
Natanzon class of potentials is invariant under this algebra, a result
different to the one obtained in SUSYQM due to the fact that the SUSYQM
operators may define a potential that does not belong to the Natanzon class %
\cite{cuper}.

\section{The $so(2,1)$ description of the \nobreak hypergeometric Natanzon
potentials}

\label{sonata} A two variable realization of the algebra $so(2,1)$ is used
with the generators taken as 
\begin{eqnarray}
{{e^{\pm i\phi }}J_{\mp }} &=&{\large {-{\frac{i\left( 1+z\right) }{2\sqrt{z}%
}}{\frac{\partial }{\partial \phi }}\mp {\frac{\sqrt{z}\left( z-1\right) }{%
z^{\prime }}}{\frac{\partial }{\partial r}}}}+  \label{generadores} \\
&&{\left( \mp \,{\frac{\sqrt{z}\left( z-1\right) z^{\prime \prime }}{%
2z^{\prime 2}}}+\,{\ \frac{\left( z-1\right) \left( \pm 1+p\right) }{2\sqrt{z%
}}}\right) }  \nonumber
\end{eqnarray}

\bigskip where $z=z(r)$ and $z^{\prime }=dz/dr$.

The expressions (\ref{generadores}) lead to the Casimir, ${%
Q=J_{0}(J_{0}+1)-J_{-}J_{+}}$, 
\begin{eqnarray}
{Q} &=&{\left( z-1\right) ^{2}\left[ {\frac{z}{z^{\prime 2}}}{\frac{\partial
^{2}}{\partial {r}^{2}}}+\,{\frac{1}{4z}}{\frac{\partial ^{2}}{\partial {%
\phi }^{2}}}+\,{\frac{ip\left( 1+z\right) }{2z\left( z-1\right) }}{\frac{%
\partial }{\partial \phi }}\right. }+  \label{casim} \\
&&{\left. \,{\frac{-{p}^{2}+1}{4z}}+\,{\frac{z{\ }z^{\prime \prime \prime }}{%
2{\ }z^{\prime 3}}}-\,{3\frac{z{\ }z^{\prime \prime 2}}{4{\ }z^{\prime 4}}}%
\right] }  \nonumber
\end{eqnarray}
The significant result that follows from (\ref{generadores}, \ref{casim}) is
the appearance of the parameter $p$ in the explicit expression for the
generators and the Casimir; this constant distinguishes a particular set of $%
so(2,1)$ generators and plays a crucial role in the $so(2,1)$ description of
the Natanzon potentials. \medskip \newline
\indent The physical problem dealt with is the derivation of the discrete
spectrum of the Hamiltonian and to this end the compact operator ${J_{0}}$
is diagonalized; the considered irreps of $so(2,1)$ are unitary and
therefore infinite dimensional and of these it is relevant the one bounded
below (the ${D^{+}}$ representation). In this representation the eigenvalue $%
m$ of the compact operator is given in terms of the eigenvalue $q$ of the
Casimir and the counter $\nu =0,1,\ldots $ as 
\begin{equation}
{m(\nu )}=\nu +\frac{1}{2}+\frac{1}{2}\,\sqrt{4\,{q}+1}  \label{eme}
\end{equation}
With the above results the $so(2,1)$ description of the Schr\"{o}dinger
equation is defined by (\ref{master}), referred to as the master equation 
\begin{equation}
{G(r)\left( Q-q\right) \Psi (r,\phi )}={\left( E-H\right) \Psi (r,\phi )}
\label{master}
\end{equation}
where $H$ is the Hamiltonian and $q$, $E$ the eigenvalues of the Casimir and
Hamiltonian, respectively.\ The function $G(r)$ ensures that the coefficient
of the second derivative of $\Psi (r,\phi )$ are the same on both sides.
From (\ref{master}) it follows that in general $q$ could be a function of $%
\nu $ which also labels $E$; in spite of the fact that $p$ does not appear
explicitly in (\ref{master}) it could also depend on $\nu $. Each
eigenfunction of the Casimir (equivalently of the Hamiltonian) has the form 
\begin{equation}
{\Psi (r,\phi )}={exp(im(\nu )\phi )g}  \label{autof}
\end{equation}
and is also an eigenfunction of the compact operator ${J_{0}}$. The function 
$g=g(r)$ is determined solving the master equation. \medskip \newline
\indent The set of Natanzon hypergeometric potentials \cite{nata} is given
by 
\begin{eqnarray}
{V_{N}} &=&{{\frac{f\,z^{2}-\left( h_{0}-h_{1}+f\right) z+h_{0}+1}{R}}}
\label{nata} \\
&+&{\left[ a+{\frac{a+\left( c_{1}-c_{0}\right) \left( 2\,z-1\right) }{%
z\left( z-1\right) }}-\,{\frac{5\Delta }{4R}}\right] \left[ \frac{z\left(
1-z\right) }{R}\right] ^{2}}  \nonumber
\end{eqnarray}
where $({a,c_{0},c_{1},f,h_{0},h_{1}})$ are the Natanzon parameters and $z$
is a solution of the differential equation 
\begin{equation}
{z(r)}^{\prime }={2\,{\frac{z\left( 1-z\right) }{\sqrt{R}}}};  \label{zeta}
\end{equation}
the other symbols that appear in (\ref{nata}, \ref{zeta}) are given by 
\begin{equation}
\tau ={c_{1}-c_{0}-a},\ \;\;\Delta ={\tau ^{2}-4\,a\,c_{0}},\ \;\;{R}={%
a\,z^{2}+\tau z+c_{0}}
\end{equation}
The $so(2,1)$ description of ${V_{N}(r)}$ is obtained after the explicit
expression for the Casimir (\ref{casim}) is put into a form similar to (\ref%
{nata}) after use of (\ref{zeta}) and the coefficients of the powers of $z$
compared; this leads to 
\begin{eqnarray}
{p(\nu )+m(\nu )}=\sqrt{{-aE(\nu )+f+1}} &=&\alpha (\nu )  \nonumber \\
{p(\nu )-m(\nu )}=\sqrt{{-c}_{0}{E(\nu )+h}_{0}{+1}} &=&\beta (\nu )
\label{relgrup} \\
\sqrt{4\,{q(\nu )}+1}=\sqrt{{-c}_{1}{\,E(\nu )+h}_{1}+1} &=&\delta (\nu ). 
\nonumber
\end{eqnarray}
From (\ref{eme}, \ref{relgrup}) ${p(\nu ),q(\nu ),m(\nu )}$ -called the
group parameters- and ${E(\nu )}$ are determined for each value of $\nu $
thus fixing a particular $so(2,1)$ irrep and the energy eigenvalue. This
implies that each eigenfunction belongs to the carrier space of an $\
so(2,1) $ which depends on $p$ see (2) and hence may depend on $\nu $ as
follows from (9). From now on $p(\nu )$, $q(\nu )$ and $m(\nu )$ are written 
$p$, $q$ and $m$. The carrier space of a specific irrep is given by 
\begin{equation}
{\Psi }_{pqm}{(r,\phi )=exp}(im\phi ){\Phi _{pqm}(r)}
\end{equation}
where \begin{flushleft}
${ \Phi_{pqm}(r)}=$
\end{flushleft}
\begin{equation}
{Kz^{\,\beta (\nu )/2}\left( 1-z\right) ^{\,\delta (\nu )/2}R^{1/4}\mbox{}%
_{2}F_{1}(-\nu ,\alpha (\nu )-\nu ;1+\beta (\nu );z)}.  \label{onda}
\end{equation}
where $K$ is a normalization constant. The energy spectrum is obtained from 
\begin{equation}
\alpha (\nu )-\beta (\nu )-\delta (\nu )=2\nu +1.  \label{abd}
\end{equation}

\section{Satellite potentials}

\label{generalres} In this Section the possibilities allowed by (\ref%
{relgrup}) for fixed values of the group parameters ($p,q$) are presented.
The study of this situation leads to the construction of the satellite
potentials. The action of the $so(2,1)$ generators on (\ref{onda}) is the
following 
\begin{eqnarray}
{J_{-}\Psi _{pqm}} &=&{{\frac{\nu \,\left( \alpha (\nu )-\nu -1-\beta (\nu
)\right) }{1+\beta (\nu )}}}{\Psi _{{pqm-1}}}  \label{jotas} \\
{J_{+}\Psi _{pqm}} &=&{-\beta (\nu )\Psi _{pqm+1}}  \nonumber
\end{eqnarray}
which follow from \cite{tabla} 
\begin{flushleft}
${\rm \mbox{}_{2}F_{1}(a+1,b+1,c+1;z)}= $
\end{flushleft}
\begin{equation}
-{\frac{1}{{zb(z-1)a}}}{\left[ c(c-1)\mbox{}_{2}F_{1}(a-1,b,c-1;z)-c(zb-c+1)%
\mbox{}_{2}F_{1}(a,b,c;z)\right] }
\end{equation}
and 
\begin{flushleft}
${\rm \mbox{}_{2}F_{1}(a-1,b,c-1;z)}=$
\end{flushleft}
\begin{equation}
-{\frac{1}{{(c-1)}}}{\left[ (b-c+1)\mbox{}_{2}F_{1}(a,b,c;z)-b(z-1)\mbox{}%
_{2}F_{1}(a,b+1,c;z)\right] }
\end{equation}
\indent It follows from (\ref{jotas}) that the action of the $so(2,1)$
generators on an eigenfunction of a given Natanzon potential gives an
eigenfuncion of a different Natanzon potential . In fact, from (\ref{jotas})
the same $z$ is present so that ${a,c_{0},c_{1}}$ are unchanged. The other
three Natanzon parameters ${f_{1},h_{01},h_{11}}$ depend on ${\nu }$ after (%
\ref{jotas}) and (\ref{relgrup}) are determined by 
\begin{equation}
\alpha _{1}(\nu \pm 1)=\alpha (\nu )\pm 1,\beta _{1}(\nu \pm 1)=\beta (\nu
)\mp 1,\delta _{1}(\nu \pm 1)=\delta (\nu )  \label{newalfas}
\end{equation}
where ${\alpha _{1}(\nu )},{\beta _{1}(\nu )},{\delta _{1}(\nu )}$ follow
from (\ref{relgrup}) by ${f\rightarrow f_{1},h_{0}\rightarrow
h_{01},h_{1}\rightarrow h_{11}}$. Thus the complete set of Natanzon
parameters is determined for the new potential which shows that the Natanzon
class is invariant under the action of the $so(2,1)$ generators.
Nevertheless the energy eigenvalue $E_{1}(\nu )$ is undetermined since (\ref%
{abd}) is a linear combination of (\ref{relgrup}). The fact that one of the
four quantities ($f$, $h_{0}$, $h_{1}$, $E(\nu )$) is arbitrary can be used
to fix the energy of the ground state.

\indent The potentials to which the eigenfunctions with ${m\pm 1}$ are
associated will be called satellite to the one related to the eigenfunction
labeled by ${m}$. The common feature to all these potentials is that the
values of both ${p}$ and ${q}$ remain unchanged. It is natural to ask, how
many satellite potentials are associated to a given one? Different
eigenfunctions of a given Natanzon potential do not necessarily belong to
the same $so(2,1)$ irrep and therefore for each such function a certain
number of satellite potentials is determined. Fix the attention in one
eigenfunction. It belongs at the same time to the set of eigenfunctions of ${%
V_{N}}$ and to the $so(2,1)$ irrep. Recall that $\nu $ labels the place of ${%
\Phi _{pqm}(r)}$ on the irrep and call $\lambda $ the label of its place in
the set of eigenfunctions of ${V_{N}}$. The above description assigns a set
of values of $(p,q,m)$ in each case and both sets must coincide, that is 
\begin{equation}
{p(\nu )=p(\lambda )},{q(\nu )=q(\lambda )},{m(\nu )=m(\lambda )}
\end{equation}
which imply 
\begin{equation}
\lambda =\nu
\end{equation}
so that the numerical value of both labels coincide on the irrep. Therefore,
the action of ${J_{+}}$ increase both $\nu $ and $\lambda $ and if there is
a maximum value of $\lambda $ for a given potential a finite number of
satellite potentials will be constructed. The maximum value $\lambda $
depends on whether the ${r\rightarrow \infty }$ limit is finite or not for
each of the satellite potentials. \medskip \newline
\indent To complete the answer consider a Natanzon potential ${V_{N}}$ one
of whose eigenfunctions, ${\Psi _{\lambda _{0}=\nu _{0}}^{\nu _{0}}}$, is in
the $(p,q)$ irrep; its place in the set of eigenfunctions is $\lambda
=\lambda _{0}$ and the one in the irrep is $\nu =\nu _{0}$. The reason for
introducing both $\nu $ and $\lambda $ (which may seem redundant because for
this particular function their numerical values coincide) is that, as shown
above, the Natanzon parameters of the satellite to ${V_{N}}$ depend on those
for ${V_{N}}$ in a way determined by (\ref{newalfas}); therefore, these
parameters may (and in fact will) be functions of $\nu $. Thus, displacement
along the irrep produces sets of Natanzon parameters that change with $\nu $
in such a way that their numerical values coincide with those for ${V_{N}}$
when $\nu =\nu _{0}$ and those for the satellite when $\nu =\nu _{0}\pm 1$.
Moving from one eigenfunction of ${V_{N}}$ to another keeps $\nu _{0}$ fixed
since in this case the Natanzon parameters do not change. The question now
is the following: if ${\Psi _{\lambda _{0}=\nu _{0}}^{\nu _{0}}}$ and ${\Psi
_{\lambda _{0}+1=\nu _{0}+1}^{\nu _{0}+1}}$ are in the same irrep, does the
same occurs for ${\Psi _{\lambda _{0}+1}^{\nu _{0}}}$ and ${\Psi _{\lambda
_{0}+2}^{\nu _{0}+1}}$? To answer this question the system (\ref{relgrup}, %
\ref{eme}) is studied taking the Natanzon parameters for $\nu =\nu _{0}$ and 
$\nu =\nu _{0}+1$, replacing $\nu $ by $\lambda $ and comparing the values
of $p$ and $q$ obtained for $\lambda =\lambda _{0}+1$ (for $\nu _{0}$) and $%
\lambda =\lambda _{0}+2$ (for $\nu _{0}+1$). It turns out that in general
these values are not the same, the analysis described above has to be
repeated for each eigenfunction and a new set of satellite potentials is
thus determined. \newline
\smallskip

\section{Particular cases}

The results that follow include the set of Natanzon parameters for the
potential and its supersymmetric partner and the rule that generates the
parameters for the satellite potentials. It has been found that for the six
cases studied the satellite potentials do not coincide with the SUSY
partner. A detailed study along the lines presented in this paper has been
performed for the Eckart potential in \cite{cosa}; where it is shown that
the satellite potential does not coincide with the supersymmetric
partner.\medskip \newline
\noindent \textbf{1.} The P\"{o}schl-Teller II potential in the notation of %
\cite{dabro} is 
\begin{equation}
{V_{PT2}}={\left( A-B\right) ^{2}-A\left( A+\alpha \right) {\ sech}(\alpha
\,r)^{2}+B\left( B-\alpha \right) {\ csch}(\alpha \,r)^{2}}.
\end{equation}
The Natanzon parameters 
\begin{eqnarray}
{a} &=&0,{{\ c_{0}}=0},{{\ c_{1}}={\alpha }^{-2}},  \nonumber \\
{f} &=&{\,{\frac{\left( 2\,A-\alpha \right) \left( 2\,A+3\,\alpha \right) }{{%
4\alpha }^{2}}}},  \nonumber \\
{{\ h_{0}}} &=&{\,{\frac{\left( 2\,B+\alpha \right) \left( 2\,B-3\,\alpha
\right) }{{4\alpha }^{2}}}}, \\
{{\ h_{1}}} &=&{{\frac{\left( A-B+\alpha \right) \left( A-B-\alpha \right) }{%
{\alpha }^{2}}}}  \nonumber
\end{eqnarray}
reproduce ${V_{PT2}}$ with ${z=tanh(\alpha \,r)^{2}}$ replaced in (\ref{nata}%
). From (\ref{relgrup}) the group parameters are 
\begin{equation}
{p=\,{\frac{A+B}{2\alpha }},\;q=\,{\frac{\left( A-B-2\,\nu \,\alpha \right)
^{2}-{\alpha }^{2}}{{4\alpha }^{2}}},\;m=\,{\frac{A-B+\alpha }{2\alpha }}}.
\end{equation}
and again from (\ref{relgrup}) 
\begin{equation}
{{\alpha }(\nu )=\,{\frac{\alpha +2\,A}{2\alpha }},\;{\beta }(\nu )=-\,{%
\frac{-2\,B+\alpha }{2\alpha }},\;{\delta }(\nu )=-{\frac{2\,\nu \,\alpha
-A+B}{\alpha }}}  \label{alfapt2}
\end{equation}
The energy spectrum is obtained from (\ref{abd}) and (\ref{alfapt2}) as 
\begin{equation}
{E(\nu )=-4\,\alpha \,\nu \,\left( \nu \,\alpha -A+B\right) }
\end{equation}
After use of (\ref{newalfas}), the action of ${J_{+}}$ leads to 
\begin{equation}
{\alpha _{1}(\nu +1)}=\,{\frac{2\,A+3\,\alpha }{2\alpha }},\;{\beta _{1}(\nu
+1)}=\,{\frac{2\,B-3\,\alpha }{2\alpha }},\;{\delta _{1}(\nu +1)}=2\,\nu +{%
\frac{B-A}{\alpha }};  \label{newpt2}
\end{equation}
calling $({A_{S},B_{S}})$ the parameters of the potential in (\ref{alfapt2})
and equate the result with the values in (\ref{newpt2}) to obtain 
\begin{equation}
{A_{S}=A+\alpha },\;{B_{S}=B-\alpha }  \label{abes}
\end{equation}
The energy spectrum for the satellite potential follows from (\ref{abd}) 
\begin{equation}
{E_{S}(\nu +1)=E(\nu )+{h_{1S}}\,{\alpha }^{2}+{\alpha }^{2}+2\,AB-{B}^{2}-{A%
}^{2}}
\end{equation}
with ${h_{1S}}$ arbitrary; if the condition that the energy vanishes for $%
\nu =0$ is imposed, then 
\begin{equation}
{h_{1S}={\frac{\left( -A+B-\alpha \right) \left( \alpha +B-A\right) }{{%
\alpha }^{2}}}}.  \label{ache1}
\end{equation}
which leads to ${E_{S}(\nu +1)=E(\nu )}$. Notice that, from (\ref{abes}), it
follows that the change in the parameters does not coincide with the one
found in SUSYQM to relate a potential with its supersymmetrtic partner \cite%
{dabro}. The chain of satellite potentials is, thus, different from the
chain of supersymmetric partners. Since the choice (\ref{ache1})
produces equal values of the energy for both the potential and it satellite
it is natural to compare this result with the one obtained in the potential
algebra of \cite{al} \ which consider the case $B=0$ and no constant term.
The change (\ref{abes}) take the value of $B$ to $-\alpha $ and therefore to
a completely different P\"{o}schl-Teller potential; the conclusion is that
the set of satellite potentials is different to the set of potentials
obtained with the potential algebra. \medskip\\
\noindent \textbf{2.} The Rosen-Morse potential 
\begin{equation}
{V_{RM}={A}^{2}+{\frac{{B}^{2}}{{A}^{2}}}+2\,B\ }tanh{(\alpha \,r)-A\left(
A+\alpha \right) {\ sech}(\alpha \,r)^{2}}
\end{equation}
is obtained from (\ref{nata}) with the Natanzon parameters 
\begin{eqnarray}
{a} &=&0,\;{c_{0}=c_{1}=1/\alpha ^{2}}  \nonumber \\
{f} &=&{4\,{\frac{A\left( A+\alpha \right) }{{\alpha }^{2}}}},  \nonumber \\
{h_{0}} &=&{{\frac{\left( -B+A\alpha +{A}^{2}\right) \left( -B-A\alpha +{A}%
^{2}\right) }{{\alpha }^{2}{A}^{2}}}}, \\
{h_{1}} &=&{{\frac{\left( B+A\alpha +{A}^{2}\right) \left( B-A\alpha +{A}%
^{2}\right) }{{\alpha }^{2}{A}^{2}}}}  \nonumber
\end{eqnarray}
with ${z=1/2+tanh(\alpha r)/2}$. The group parameters are found to be 
\begin{eqnarray}
{p} &=&\frac{2A}{\alpha }-m+1,  \nonumber \\
{q} &=&{\left( \nu +1-m\right) \left( \nu -m\right) },  \label{gruprm} \\
{m} &=&{\,{\frac{-{A}^{2}+\nu \,{\alpha }^{2}-A\alpha -B+{\nu }^{2}{\alpha }%
^{2}}{2\left( \nu \,\alpha -A\right) \alpha }}}  \nonumber
\end{eqnarray}
From (\ref{relgrup}, \ref{gruprm}) it follows 
\begin{eqnarray}
{\alpha (\nu )} &=&{\frac{2\,A+\alpha }{\alpha }},  \nonumber \\
{\beta (\nu )} &=&{{\frac{{\nu }^{2}{\alpha }^{2}-2\,\nu \,\alpha \,A+{A}%
^{2}-B}{\left( A-\nu \,\alpha \right) \alpha }}}, \\
{\delta (\nu )} &=&{{\frac{{A}^{2}+B+{\nu }^{2}{\alpha }^{2}-2\,\nu \,\alpha
\,A}{\left( A-\nu \,\alpha \right) \alpha }}}  \nonumber
\end{eqnarray}
and from (\ref{abd}) the energy spectrum is given by 
\begin{equation}
{E(\nu )={A}^{2}-\left( A-\nu \,\alpha \right) ^{2}+{\frac{{B}^{2}}{{A}^{2}}}%
-{\ \frac{{B}^{2}}{\left( A-\nu \,\alpha \right) ^{2}}}}
\end{equation}
For the satellite potential it is found from (\ref{abd}) 
\begin{eqnarray}
{{\alpha _{1}(\nu +1)}} &=&{2\,{\frac{A+\alpha }{\alpha }}},  \nonumber \\
{{\ \beta _{1}(\nu +1)}} &=&{-{\frac{A-\nu \,\alpha }{\alpha }}-1+{\frac{B}{%
\left( A-\nu \,\alpha \right) \alpha }}}, \\
{{\delta _{1}(\nu +1)}} &=&{-{\frac{A-\nu \,\alpha }{\alpha }}-{\frac{B}{%
\left( A-\nu \,\alpha \right) \alpha }}}  \nonumber
\end{eqnarray}
Setting (${A_{S},B_{S},p_{S},q_{S}}$) instead of ${(A,B,p,q)}$ in (\ref%
{gruprm}) with $\nu \rightarrow \nu +1$ and requiring that ${p(\nu
)=p_{S}(\nu +1),q(\nu )=q_{S}(\nu +1)}$ it is found 
\begin{equation}
{{\ A_{S}}=A+\,\alpha /2,{\ \;B_{S}}=-\,{\frac{\left( 2\,\nu \,\alpha
+\alpha -2\,A\right) \left( \nu \,{\alpha }^{2}-A\alpha -2\,B\right) }{4\nu
\,\alpha -A}}}
\end{equation}
The energy of the satellite is found to be 
\begin{equation}
{E_{S}(\nu +1)=E(\nu )+\left( {h_{1S}-h_{1}}\right) \,{\alpha }^{2}}
\end{equation}
with obvious meaning for ${h_{1S}}$. The same comment made for the previous
example are valid here. \medskip \newline
\indent The conclusion that follows from the above examples is that the
chain of potentials defined by the action of the $so(2,1)$ generators is
different from the one defined by the action of the SUSYQM operators and of
the potential algebra.

\bigskip


\begin{thebibliography}{99}
\bibitem{corsa}  Cordero P and Salam\'{o} S 1993 \textit{Found. Phys.} \textbf{23} 675 ; 1994 \textit{J. Math. Phys.} \textbf{35} 3301   

\bibitem{nata} Natanzon G A 1979 \textit{Teor. Mat. Fiz.} \textbf{38} 146  

\bibitem{mesa} del Sol Mesa A, Quesne C and  Smirnov Y F 1998 \textit{J.
Phys. A: Math. Gen.} \textbf{31} 321

\bibitem{cuper1} For a complete review of SUSYQM see Cooper F, Khare A and Sukhatme U 1995 \textit{Phys. Rep.} \textbf{251}  2671

\bibitem{gede} Gendenshtein L 1983 \textit{JETP Lett.} \textbf{38} 356

\bibitem{cosa} Codriansky S, Cordero P and Salam\'{o} 1999 S \textit{J. Phys.
A: Math. Gen.} \textbf{32} 6287

\bibitem{tabla} Gradshteyn I S and Ryzhik I M 1965 \textit{Table of
Integrals, Series and Products} (New York: Academic Press)

\bibitem{cuper} Cooper F, Ginocchio J N and Khare A 1987 \textit{Phys. Rev. D%
} \textbf{36} 2458

\bibitem{dabro} Dabrowska J, Khare A and Sukhatme A P 1988 \textit{J. Phys.
A: Math. Gen.} \textbf{21} L195

\bibitem{al} Alhassid Y, G\"{u}rsey F and Iachello F 1986 \textit{Ann Phys}%
\textbf{\ 167} 181
\end{thebibliography}
\end{document}